\documentclass[12pt,eqsecnum]{article}
\usepackage{amsmath,amssymb,graphicx,mathrsfs} 

\usepackage{psfrag}
\usepackage{subfigure}
\usepackage{tabularx}

\parskip=\baselineskip
\renewcommand{\baselinestretch}{1.2}

\newcommand\cc[1]{#1^{^{\kern-6pt \circ}}\kern2pt}

\newcommand{\dd}{{\rm d}}

\newcommand{\m}{\mu}
\newcommand{\n}{\nu}

\def\be{\begin{equation}}
\def\ee{\end{equation}}
\def\bea{\begin{eqnarray}}
\def\eea{\end{eqnarray}}
\def\ba{\begin{array}}
\def\ea{\end{array}}
\def\bi{\begin{itemize}}
\def\ei{\end{itemize}}

\newcommand{\beq}{\begin{equation}}
\newcommand{\eeq}{\end{equation}}
\newcommand{\beqn}{\begin{eqnarray}}
\newcommand{\eeqn}{\end{eqnarray}}
\newcommand{\bga}{\begin{align}}

\def\dalemb#1#2{{\vbox{\hrule height .#2pt
\hbox{\vrule width.#2pt height#1pt \kern#1pt
\vrule width.#2pt}
\hrule height.#2pt}}}

\begin{document}

\def\thataddress{\small Department of Physics, University of Crete, Heraklion 71003, Greece}
\renewcommand\author[1]{#1}

\begin{center}
{\Large {\bf  Stochastic Quantization and AdS/CFT }}
\end{center}
%\centerline{[Draft Version of \today]}
\vspace{.5cm}
\centerline{\bf 
\author{Diego S. Mansi\footnote{{\small \tt diego.mansi@mi.infn.it}} and  Andrea Mauri\footnote{{\small\tt andrea.mauri@mi.infn.it}}}} 
\vspace{.3cm}
\centerline{\small \it
Dipartimento di Fisica Teorica, Universita degli Studi di Milano}
\centerline{\small \it Via Celoria, Milano 20133, Italy}
\vspace{.3cm}
\centerline{\bf 
\author{Anastasios C. Petkou\footnote{{\small\tt petkou@physics.uoc.gr}}}}
\centerline{\it\thataddress}

%\keywords{}
%\preprint{}

%\vspace{2cm}

\begin{abstract}
We argue that there is a relationship between stochastic quantization and AdS/CFT, and we present an explicit calculation to support our claim. In particular, we show that a  conformally coupled scalar with 
$\phi^4$ interaction on AdS$_4$ is related, via stochastic quantization as well as via AdS/CFT, to a massless scalar with $\phi^6$ interaction in 3d. We show that our results have an underlying geometric origin, which might help to elucidate further the proposed relationship between stochastic quantization and holography. 
\end{abstract}

%\tableofcontents

%\pagebreak

%\parskip= 2pt
%\renewcommand{\baselinestretch}{.2}
%\tableofcontents

%\parskip=10pt
\renewcommand{\baselinestretch}{1.2}

\section{Introduction}
The possibility that stochastic quantization is related to AdS/CFT has been discussed before (e.g. \cite{DPolyakov,Periwal}), however the discussion has not picked up momentum mainly due to the absence of an explicit example. It is not hard to anticipate such a relationship. In the stochastic quantization scheme of Parisi \& Wu (\cite{ParisiWu}) the correlation functions of an Euclidean $d$-dimensional field theory arise as equilibrium configurations, for large ``fictitious'' times, of the corresponding correlation functions of a $d+1$-dimensional field theory described by a Fokker-Planck action. On the other hand, in AdS/CFT the generating functional for connected correlation functions of a $d$-dimensional field theory arises as the appropriately renormalized on-shell action of a $d+1$-dimensional gravitational theory. Therefore, a connection could be established if the stochastically quantized action is somehow related to the boundary action of AdS/CFT and if the Fokker-Planck and the holographic bulk actions are also related. Clearly, we also need to relate the stochastic "time" to the holographic direction.\footnote{Related ideas have recently appeared in the lattice approach to quantum gravity \cite{AmbjornLoll}.}  Then, such a relationship would imply a  profound connection between stochastic processes and gravitation.  

 In this work we revisit the idea that stochastic quantization is
 related to AdS/CFT and we present an explicit example to support it. We start by
 sketching a simple formal correspondence between stochastic
 quantization and AdS/CFT. Namely, we show that the partition function
 of stochastic quantization corresponds to an average over
 holographic partition functions, if we identify the Fokker-Planck
 and the bulk actions, and also the initial classical action with the
 holographic boundary effective action. Our explicit example involves
 a conformally coupled scalar with $\phi^4$ interaction in fixed
 AdS$_4$. We show that the leading terms in a large coupling expansion of the holographic effective action of the model,
 give the 3d action of a massless scalar with $\phi^6$ interaction written, curiously, in an unconventional manner.  Next, starting
 from the latter 3d action  we use stochastic quantization to arrive
 at its corresponding 4d Fokker-Planck action. The leading terms in a large coupling expansion of that Fokker-Planck action give precisely (i.e. including the numerical coefficients),  the initial 4d action of a massless scalar with $\phi^4$ interaction. The latter action is actually  equivalent to that of a conformally coupled scalar with $\phi^4$ interaction on
 AdS$_4$. Hence, our explicit example demonstrates that the above 3d and 4d field theories are related {\it both} via AdS/CFT  as well as via stochastic quantization. We consider our results as a strong indication that stochastic quantization and AdS/CFT are intimately related.  We then discuss the general conditions under which such a relationship might arise focusing on the role of boundary conditions.  
 Finally,
 we point out that our results above have a geometric origin. Indeed, both the 4d and 3d actions that are involved in our
 example are merely disguised 4d and 3d gravitational actions
 for conformally flat metrics. Then, the boundary condition that enables the calculation of the boundary effective action in AdS/CFT arises as the stationarity condition for a system involving bulk and boundary gravity.  This observation might help to elucidate further the  relationship between stochastic quantization and AdS/CFT in our specific example. 
 
 \section{Is Stochastic Quantization related to AdS/CFT?}

It is not hard to sketch a formal connection between stochastic quantization (see Appendix A for a condensed review) and AdS/CFT. The Boltzmann weight for a $d$-dimensional Euclidean theory of the scalar field $\phi(\vec{x})$ is (we set henceforth $\hbar=1$)
\beq
\label{Boltzmann}
{\cal P}[\phi]\equiv\frac{1}{{\cal Z}_d}e^{-S_{\rm cl}[\phi]}\,,\quad{\rm with}\quad {\cal Z}_d=\int[{\cal D}\phi]\,e^{-S_{\rm cl}[\phi]}\,,
\eeq
or equivalently $\int[{\cal D}\phi]\,{\cal P}[\phi]=1$. The extended scalar field $\phi(\vec{x})\mapsto \phi(t,\vec{x})$ satisfies the Langevin equation
\beq
\label{Langevin}
\frac{\partial\phi(t,\vec{x})}{\partial t}+ \kappa \frac{\delta S_{\rm cl}[\phi]}{\delta\phi(t,\vec{x})} =\eta(t,\vec{x})\,,
\eeq
where $\kappa$ is a generic kernel. The source $\eta(t,\vec{x})$ is a ``white noise'' defined by the following partition function and correlation functions
\bea
\label{Part_eta}
&&{\cal Z} =\int [{\cal D}\eta]\exp\left[-\frac{1}{4\kappa}\int_{-T}^{0}\dd t\int \dd^{d}x\,\eta^2(t,\vec{x})\right]\,,\\
&&\langle\eta(t,\vec{x})\rangle =0\,,\\
&&\langle \eta(t_1,\vec{x}_1)\eta(t_2,\vec{x}_2)\rangle = 2\kappa\delta^d(\vec{x}_1-\vec{x}_2)\delta(t_1-t_2)\,.
\eea
To make the connection with AdS/CFT we need to depart slightly from the standard stochastic quantization procedure where the fictitious "time" interval is taken to be $t\in [0,T]$. In that case, one fixes the initial field configurations at $t=0$ and lets the system evolve in $t$. The crucial point is then \cite{ParisiWu} that at $T\rightarrow\infty$ the fields and their equal "time" correlations functions relax to their equilibrium values - the latter being identified with properly quantized configurations.    

Here instead we take the fictitious "time" interval to be $t\in[-T,0]$ such that starting from any finite initial  $t=-T$, the fields evolve via the Langevin towards their values at $t=0$ which we denote as $\phi_{-T}(0,\vec{x})$. Sending then the initial "time" $-T\rightarrow -\infty$, the system reaches thermal equilibrium at $t=0$ i.e. the field configurations at $t=0$  are properly quantized.  Accordingly, we have to impose an initial  distribution for the field $\phi(t,\vec{x})$ at $t=-T$ by means e.g. of a delta function as
\beq
\label{deltaBC}
P_{t=-T}[\phi]=\Pi_x\left\{\delta^d\left[\phi(-T,\vec{x})\right]\right\}\,.
\eeq
Here we have chosen a vanishing initial configuration having in mind to take a large T limit. The t=0  correlation functions, which are evaluated as stochastic averages over the white noise with Boltzmann weight that of (\ref{Part_eta}),  relax into those of the $d$-dimensional theory (\ref{Boltzmann}), namely
\beq
\label{correlT}
\lim_{T\rightarrow\infty}\langle\phi_{-T}(0,\vec{x}_1)\phi_{-T}(0,\vec{x}_2)...\phi_{-T}(0,\vec{x}_n)\rangle_{\eta} =\langle\phi(\vec{x}_1)\phi(\vec{x}_2)...\phi(\vec{x}_n)\rangle_{S_{cl}}\,.
\eeq
One can turn the stochastic averages over the white noise into "path integrals" over the scalar fields changing variables $\eta\mapsto\phi$. After a straightforward calculation (see e.g. \cite{Damgaard}) and with our choice of initial data we get
\beq
\label{Zstoch}
{\cal Z} =\int [{\cal D}\phi(0)]e^{-S_{\rm cl}[\phi(0)]/2} \int[{\cal D}\phi]e^{-{\cal S}_{FP}}\,,
\eeq
where the Fokker-Planck action ${\cal S}_{FP}$ and the "path integral" measure are
\begin{align}
\label{Sfp}
{\cal S}_{FP} &= \int_{-T}^0 dt\int d^d\vec{x}\,\left[\frac{1}{4\kappa}\dot{\phi}^2+\frac{\kappa}{4}\left(\frac{\delta S_{\rm cl}}{\delta\phi}\right)^2-\frac{\kappa}{2}\frac{\delta^2S_{\rm cl}}{\delta\phi^2}\right]\,,\\
\label{tildeDphi}
[{\cal D}\phi]&= \prod_{-T<t<0}[{\cal D}\phi(t)]\,.
\end{align}
To obtain (\ref{Zstoch}), the following (formal) result\footnote{The determinant gives rise generically to infinities of the form $\delta^d(0)$ that - when properly regularized - act as counterterms to some of the divergences that arise in the perturbative expansion  \cite{Zinn-Justin}.} for the determinant was used
\beq
\label{detEtaPhi}
\det\left(\frac{\delta\eta}{\delta\phi}\right) =\exp\frac{\kappa}{2}\left[\int_{-T}^0dt\int d^d\vec{x}\,\,\frac{\delta^2 S_{cl}[\phi]}{\delta\phi(t,\vec{x})^2}\right]\,.
\eeq
Notice that starting with a conventional scalar theory $S_{cl}$, the Fokker-Planck action ${\cal S}_{FP}$ is generically non-relativistic. A recent discussion on this property of stochastic quantization has appeared in \cite{Horava}. 

Consider next a $d+1$-dimensional scalar theory on an fixed
asymptotically AdS background. The implementation of holography in
such a simple case requires the evaluation of the following partition
function using a semiclassical approximation as \beq
\label{Zholo}
{\cal Z}_{\rm hol}[\phi_0]=\int [{\cal
    D}\phi]_{\phi_0}e^{-S_{d+1}[\phi]}\equiv e^{W_d[\phi_0]}\,, \eeq
where $[{\cal D}\phi_0]_{\phi_0}$ denotes path integration with
Dirichlet boundary conditions $\phi|_{\partial{\cal
    M}}=\phi_0$. Barring important regularization questions, which are
however well understood, the path integral yields the generating
functional $W_d[\phi_0]$ of connected correlation functions of a
composite operator ${\cal O}$ with dimension $\Delta$ in a
$d$-dimensional (generically conformal) field theory.

The arbitrary boundary conditions $\phi_0(\vec{x})$ are external
sources for the operator ${\cal O}$. It is important that the scaling
dimension of ${\cal O}$ is above the unitarity bound of a
$d$-dimensional CFT, namely $\Delta >d/2-1$.  Hence, a path integral
over $\phi_0$, corresponding to the quantization of $\phi_0$, will
generically produce inconsistencies such as negative probabilities or
negative norm states. Nevertheless, there are known cases where
$\phi_0$ is a normalizable mode as well and hence it can correspond to
an operator $\tilde{\cal O}$ with dimension $\tilde{\Delta}=d-\Delta
>d/2-1$. A well-known example is the conformally coupled scalar field
in 4-dimensions. In such cases the Euclidean functional
$W_{d}[\phi_0]$ itself can be used to construct a well defined
Boltzmann weight for a $d$-dimensional theory. In other words,
$W_d[\phi_0]$ can be interpreted as an {\it effective action} i.e. we can write $W_d[\phi_0]\equiv \Gamma_d[\phi_0]$. Now, the leading term of the effective action $\Gamma_d[\phi_0]$ is a classical action that we denote as $I_d[\phi_0]$. Then, it is natural to take a further
step and define (see also \cite{Marolf}) \beq
\label{Z'holo}
{\cal Z}'=\int [{\cal D}\phi_0]e^{-I_d[\phi_0]}{\cal Z}_{\rm hol}[\phi_{0}]=\int [{\cal D}\phi_0]e^{-I_d[\phi_0]}\int [{\cal D}\phi]_{\phi_0}e^{-S_{d+1}[\phi]}\,,
\eeq
as an average, with weight $I_d[\phi_0]$, of the holographic partition functions. 

We notice now a strong formal similarity between
(\ref{Zstoch}) and (\ref{Z'holo}) provided we make the following
correspondences: 
\bea {\rm S. Q.}& : & {\rm AdS/CFT} \label{StoHo}  \\ {\cal
  S}_{FP}[\phi] & \leftrightarrow & S_{d+1}[\phi]\nonumber 
\\ S_{cl}[\phi_0] & \leftrightarrow & 2 I_d[\phi_0]\nonumber\\
{\rm stochastic \,\,"time"} &\leftrightarrow & {\rm holographic \,\,direction}\nonumber \,.
\eea 
At a first glance, the above formal similarity may appear too optimistic. As we have previously commented, any
conventional $d$-dimensional action $S_{cl}$ would lead to a
non-relativistic Fokker-Planck action ${\cal S}_{FP}$. This seems
irreconcilable with the standard AdS/CFT dictionary where both the
bulk $S_{d+1}$ and the boundary $I_d$ actions are
relativistic. Furthermore, the boundary action is always conformal. This means that the presumed relationship between stochastic quantization and AdS/CFT is non-generic. Indeed, the explicit example we present below is special and has a geometric origin. Nevertheless, we do believe that the relationship between stochastic quantization and AdS/CFT even for such special cases could shed light into certain quantum properties of spacetime. 

\section{From the bulk to the boundary: AdS/CFT}
To give precise meaning to the formal correspondence sketched above we
consider the model studied in \cite{dHP,dHPP} of a conformally coupled
scalar with $\phi^4$ interaction on fixed Euclidean AdS$_4$
\begin{equation}
\label{phi4}
I = \int d^4x \sqrt{g}\left(\frac{1}{2} \,g^{\mu
  \nu}\partial_{\mu}\phi \,\partial_{\nu}\phi + \frac{1}{2} m_{\ell}^{2} \phi^2 +
\frac{\lambda}{4} \phi^4 \right)\,,\,\,\, x^{\mu}=(r,\vec{x})\,,
\end{equation}
where $\ell$ is the radius of AdS, determining the mass scale
$m_{\ell}^{2}\,\ell^{2}=-2$. The dimensionless coupling $\lambda$ is kept
general. 
%In fact, Here we can neglect the pure gravity terms because we will
%eventually be interested in expanding the theory around an instanton
%configuration: the contribution to the background geometry of these
%solutions can be safely ingnored since their stress energy
%tensor is vanishing. 
Upon introducing Poincar\'e coordinates and rescaling the
field as
\begin{align}
ds^2 &= \frac{\ell^2}{r^2} (dr^2 + d\vec{x}^2)\,, &g_{\mu \nu} &=
\Omega^{-2}(x)\eta_{\mu\nu}\,, &\Omega(x) &= \frac{r}\ell\,, &\phi &=
\Omega(x) f\,,
\end{align} 
the action becomes 
\begin{equation}
\label{4daction}
I=I_f+I_{div}=\int^{\infty}_{0} dr \int d^3\vec x \left(\frac{1}{2} \,\eta^{\mu
  \nu}\partial_{\mu}f \,\partial_{\nu}f + \frac{\lambda}{4} f^4
\right)+\int d^3\vec x \left. \frac{f^{2}}{2r}\right|^{\infty}_{0}\,.
\end{equation}
The last term is divergent and needs to be renormalized by the
addition of appropriate counterterms (see e.g. \cite{Papadimitriou}).  Hence, this simple model essentially reduces to a massless theory on
"half" 4-dimensional flat space. The asymptotic boundary resides at
$r=0,\infty$ and is isomorphic to $S^3$. As usual in the AdS/CFT
\cite{wittenAdSCFT}, we remove the point at $r=\infty$
and we are left with a theory living on $\mathbb{R}^3$ (the space at
$r=0$). This is  consistent with conformal invariance.\footnote{Under
  the conformal inversion $x^\mu\mapsto \hat{I}\,x\equiv x^\mu/x^2$,
  scalar fields with dimension $\Delta>0$ behave as $\phi(x)\mapsto
  x^{-2\Delta}\phi(\hat{I}\,x)$. Hence, the finiteness of fields in
  the origin is preserved under conformal transformations if the
  fields vanish at infinity.} Physically sensible boundary conditions
must imply this {\it regularity condition}, namely the vanishing of
the fields at the "horizon" point $r=\infty$ of AdS. The reverse does
not hold in general.

 For $\lambda>0$, the equations of
motion for the action  (\ref{4daction})
\begin{equation}\label{eqmo}
-\partial^{\mu}\partial_{\mu} f+\lambda f^3=0\,,
\end{equation}
possess the following 5-parameter family of solutions with vanishing
stress tensor (hence, they remain exact solutions in the presence of
gravity \cite{dHPP})
\begin{equation}\label{instsol}
  \hat{f}(r,\vec{x})=k\,\frac{b}{-b^2+(r+r_0)^2+(\vec{x}-\vec{x}_{0})^{2}}
  \,\,,\,\,\,\,\,\,\,k=\sqrt{\frac{8}{\lambda}}\,.
\end{equation}
The istantonic nature of this type of solutions requires $\lambda$ to
be finite.  The parameter $b$ determines the instanton size, while
$(-r_0,\vec{x}_{0})$ may be viewed as the coordinates of the instanton
center.\footnote{Notice that for a solution of of (\ref{eqmo}) to
  exist for $\lambda>0$ the instanton center must lie outside the half
  $\mathbb{R}^4$.} The solution is regular for all $r>0$ when
$r_0>b>0$.

The upshot of holography is the calculation of the renormalized
on-shell action with given boundary conditions. In the present case,
the general solution of (\ref{eqmo}) behaves near $r=0$ as 
\begin{equation}\label{soleom}
f(r,\vec{x}) =\phi_0(\vec{x})+r\phi_1(\vec{x})+O(r^2)\,, 
\end{equation} where
$\phi_0(\vec{x})$ and $\phi_1(\vec{x})$ are the two arbitrary data
necessary to determine the general solution of the 2nd order
differential equation (\ref{eqmo}). It was shown in \cite{dHP, dHPP}
that evaluating (\ref{4daction}) on-shell as a functional of $\phi_0(\vec{x})$
yields the 3-dimensional effective action for a composite scalar
operator with dimension $\tilde{\Delta}=1$ as 
\begin{equation} \label{effaction}
I_f^{on-shell}[\phi_0]=-\Gamma_d[\phi_0]\,.  
\end{equation} 
To achieve that we need to impose boundary conditions relating
$\phi_1(\vec{x})$ to $\phi_0(\vec{x})$. Such conditions can be 
generically expressed as ${\cal F}(\phi_0,\phi_1)=0$ for some
functions ${\cal F}$ \cite{Papadimitriou}. There is a general method
to evaluate the boundary on-shell action with given boundary
conditions, which is essentially the Hamilton-Jacobi method for field
theory \cite{Salopek}. However, in our simple model we can
take a more direct approach making use of the exact solution
(\ref{instsol}). Consider the Hamiltonial formulation of our model which arises in a standard and simple way from the finite part in the rhs of (\ref{4daction}). 
In this case we have
\begin{equation}\label{hamaction} 
I_f = \int^{\infty}_{0} dr \int d^3\vec{x}\left[ \pi\,\partial_r f -
  \mathcal{H} \right]\,\, , \qquad  \mathcal{H} = \frac{1}{2} \left(\pi^2 -
\partial^i f \partial_if - \frac{\lambda}{2} f^4   \right)\,.
\end{equation}
Suppose now that on-shell the following condition holds
\begin{equation}
\label{hamcondgen}
{\cal H}_{on-shell}(\pi,f) = \partial_i {\cal V}_i(f)\,,
\eeq
for some functional ${\cal V}_i(f)$. This would greatly simplify the calculation of the on-shell value of (\ref{hamaction}) since, from the one hand it implies that the contribution of the Hamiltonian in the on-shell action is a total spatial derivative and hence vanishes, and from the other hand it gives an on-shell relationship between $\pi$ and $f$ such that the kinetic term in (\ref{hamaction}) can be written (at least term-by-term in some expansion) as a total $r$-derivative. Generically, the functional ${\cal V}_i(f)$ in (\ref{hamcondgen}) can be calculated in a spatial derivative expansion and the coefficients are fixed requiring consistency with the e.o.m. \cite{Salopek}.  Here, we use input from the exact solution (\ref{instsol}) on which the  Hamiltonian density is
\beq \label{HamBC} \hat{\mathcal{H}}
= \frac{1}{2}\left(\hat{\pi}^2- \partial^i\hat{f}\partial_i \hat{f}
-\frac{\lambda}{2}\hat{f}^4 \right) = -\frac{1}{6} \partial^i
\partial_i \hat{f}^2 \,,
\eeq
 where the canonical momentum is
defined by $\pi=\partial_r f$. Hence, if we are interested in calculating the on-shell action up to two spatial derivatives we could use (\ref{HamBC}) for generic solutions of the e.o.m. 

As mentioned above, (\ref{HamBC}) is at the same time a boundary condition for the generic solution (\ref{soleom}),
since it relates $\phi_1(\vec{x})$ to $\phi_0(\vec{x})$. Explicitly the latter relation yields (we use now unhatted variables denoting a general solution of the e.o.m.)
\bea \label{momon1}
\pi(r,\vec{x}) & = & \pm\sqrt{\frac{\lambda}{2} f^4 + 
  \partial^i f \partial_i f - \frac{1}{3}\partial^i
  \partial_i f^2 } = \pi_0(\vec{x}) + r \, \pi_1(\vec{x}) +
O(r^2) \,,\\ 
\label{momon2}
\pi_0(\vec{x}) & = & \phi_1(\vec{x}) = \pm\sqrt{\frac{\lambda}{2} \phi_0^4 + 
  \partial^i \phi_0 \partial_i \phi_0 - \frac{1}{3}\partial^i
  \partial_i \phi_0^2 }\,.
\eea 
An alternative but revealing way to express the boundary condition (\ref{momon2}) is
\beq
\label{momR}
\pi_0(\vec{x}) =\pm \sqrt{\frac{\lambda}{2}}\phi_0^2(\vec{x})\Biggl[1 +\frac{1}{3\lambda}{\cal R}_3\Biggl]^{1/2}=\pm\sqrt{\frac{\lambda}{2}}\phi_0^2(\vec{x})\left(1+\frac{1}{6\lambda}{\cal R}_3+O(\lambda^{-2})\right)\,,
\eeq
where
\beq
\label{R3}
{\cal R}_3\equiv {\cal R}_3\left[g=\phi^2_0\eta\right] = -\phi^{-2}_0\Bigl[2\partial_i\ln\phi_0(\vec{x})\partial_i\ln\phi_0(\vec{x})+4\partial_i\partial_i\ln\phi_0(\vec{x})\Bigl]\,,
\eeq
is the scalar curvature of a conformally flat 3-dimensional metric $g_{ij}(\vec{x})=\phi^2_0(\vec{x})\eta_{ij}$. We term (\ref{HamBC}) {\it
 Hamiltonian boundary condition} since it implies that the
Hamiltonian density retains the form it has on an exact solution
i.e. on a solution where {\it both} $\phi_0$ and $\phi_1$ are
completely fixed. Finally, we can substitute these results in
(\ref{hamaction}) and consider a large $\lambda$ expansion to obtain
\begin{equation} \label{3daction1}
I_f^{on-shell}[\phi_0]=-\Gamma[\phi_0] = - \sqrt{\frac{1}{18\lambda}} \int d^3\vec{x}
\left(\lambda \phi_0^3 + \frac{1}{\phi_0}\,\partial^i \phi_0
\partial_i \phi_0 +  \dots \right)\,,
\end{equation}
where the dots denote terms $O(1/\lambda)$. This effective action has the intriguing  property that it is a disguised form of a well-known action in 3d. Indeed, defining $\Phi^2 =
 \frac{8}{\sqrt{18 \lambda}} \phi_0$ we find from (\ref{3daction1})
\begin{equation}\label{3daction2}
  \Gamma[\phi_0] = \Gamma[\Phi] = \int d^3\vec{x} \left( \frac{1}{2} \,\partial^{i}\Phi
  \,\partial_{i}\Phi + g\,\Phi^6 + \dots \right)\,,
\end{equation}
and the coupling constant $g =\left(\frac{3\lambda}{16}\right)^{2} $. The terms shown in (\ref{3daction1}) and (\ref{3daction2}) are then the classical action of a 3d conformal theory.\\

\section{From the boundary to the bulk: stochastic quantization }

In the AdS/CFT example above the bulk theory is holographically related to a boundary action written in terms of elementary fields. The description passes through the non-canonical 3d action (\ref{3daction1}).  We will show now that  this property can be understood, in the opposite direction,  in terms of stochastic
quantization. Namely, we will associate the Fokker-Plank action to the action of the bulk theory
and the boundary effective action to  $S_{cl}$. The Hamiltonian boundary
conditions (\ref{HamBC}) will play a crucial part in this correspondence. This way we will provide an explicit realization of the formal correspondence (\ref{StoHo}) between stochastic quantization and AdS/CFT.

To implement our idea we apply the stochastic quantization procedure to the following 3d classically conformal action 
\begin{equation} \label{3dbase}
S_{cl}[\phi] = \frac{2}{\sqrt{18\lambda}}\int d^3
\vec{x}\left(\frac{1}{\phi}\partial^i\phi\partial_i\phi + \lambda
\phi^3 \right)\,.
\end{equation}
Notice that our starting action is {\it two times} the holographic effective action we found in (\ref{3daction1}), in accordance with our identification in (\ref{StoHo}). As explained at the end of the previous Section, this unconventional action is related to a canonical $\phi^6$ model by a simple field redefinition. It is then straightforward  to compute\footnote{As mentioned before, the term (\ref{detEtaPhi}) gives rise to  $\delta^d(0)$ infinities in field theory and is irrelevant here.} the 4d Fokker-Plank action inserting (\ref{3dbase}) in (\ref{Sfp}):
\begin{eqnarray}
\left(\frac{\delta S_{cl}}{\delta \phi}\right)^2 & = & \frac{2}{9} \left[ 9 \lambda \phi^4 - 12 \phi \Box \phi + 6 \partial^i\phi\partial_i\phi + \dots\right]\,,  \\
%\frac{\delta^2 S_{cl}}{\delta \phi^2} & = & \frac{2}{\sqrt{18\lambda}} \left[ 6 \lambda \phi + \dots \right] \\
S_{FP} [\phi] & = &  \int dt \int d^3\vec{x} \left[\frac{1}{2} \partial^{\mu}\phi \partial_{\mu}\phi + \frac{\lambda}{4} \phi^4 + \dots \right] \,.
\end{eqnarray} 
Here we have chosen for definiteness $\kappa=1/2$ and the dots stand for subleading terms in a large $\lambda$ limit. Hence we see that  the leading terms for large-$\lambda$ in the stochastic quantization of $S_{cl}[\phi]$ reproduce precisely (i.e. including numerical coefficients!) the bulk action (\ref{4daction}) for the conformally coupled scalar field, if we identify the stochastic "time" with the holographic direction $r$.   

Now we would like to better understand why this connection takes place trying to give a unified picture. In order to do so we consider the general Fokker-Plank system (\ref{Sfp}) in Hamiltonian formalism
\begin{equation}\label{FPHP}
I = \int_{-\infty}^{0}dt \int d^3\vec{x} \left[ \dot{\phi}\pi -
  \mathcal{H}_{FP}\right] \,\,\,\,\, {\cal H}_{FP} = \frac{1}{2} \left({\pi}^2 -
\frac{1}{4}\left(\frac {\delta S_{cl}}{\delta\phi}\right)^2 +
\frac{1}{2}\frac{\delta^2S_{cl}}{\delta\phi^2}\right)\,,
\end{equation}
where $\pi = \dot{\phi}$ is the standard canonical momentum and again we chose $\kappa=1/2$. We notice at first that the theory lives in ``half'' 4-dimensional flat space. In section 3 we have shown that the conformally coupled scalar model on $AdS_4$ can be reduced to a theory on this same space. Then the 4-dimensional space generated by the addition of the 
fictitious time direction in stochastic quantization exactly reproduces the space-time associated to our initial bulk system. More generally it is possible to think of adding suitable time direction(s) and allow for more general fictitious time evolutions to reproduce different space-time structures to be associated to other gravitational systems. Having this in mind, we would like to perform an holographic analysis of the simple FP model in (\ref{FPHP}), now considered as our initial bulk theory.\footnote{In the context of AdS/CFT it can be shown that an action such as (\ref{FPHP}) can arise considering a non-conformally coupled scalar on fixed AdS$_4$ and taking the non-relativistic limit.} Therefore we evaluate
the variation of the action on-shell:
\begin{equation}\label{vary}
\delta I^{o.s.} = - \int d^3\vec{x} \, \delta\phi_0 \,\pi_0\,.
\end{equation} 
Now we consider a solution $\phi(r,\vec{x})$ of the Fokker-Plank
equations of motion for which 
\begin{equation}\label{bound}
\pi(0,\vec{x}) =
  -\frac{1}{2} \frac {\delta S_{cl}}{\delta\phi}\,.
\end{equation} 
It is clear from (\ref{vary}) that this selects a particular class of solutions with
specific boundary conditions such that we simply have $I^{o.s.} = \frac{1}{2}
S_{cl}$. This is the identification we were looking for in (\ref{StoHo}), with the correct overall coefficient. In this case the holographic generating functional of connected diagrams for the boundary theory is
directly related to the stochastichally quantized action. In our scalar model, the generating functional can  also be interpreted as an effective action. 
In any case, it is clear that the specific boundary condition (\ref{bound}) is the key element to obtain an exact correspondence between the action $S_{cl}$ and the boundary effective action of holography.

It is easy to see that the Hamiltonian boundary condition (\ref{HamBC}) we had to introduce in our scalar example exactly coincides with (\ref{bound}).
In fact we have 
\begin{equation}
\frac{1}{4}\left(\frac {\delta S_{cl}}{\delta\phi}\right)^2 =
\frac{1}{18 \lambda}\left( 6 \lambda \partial^i\phi\partial_i\phi -12\lambda
\phi \Box \phi + 9 \lambda^2 \phi^4 \right. + \dots) \equiv
\pi(0,\vec{x})^2 =  \dot{\phi}^2\,,
\end{equation}
and again we had to consider a large $\lambda$ limit.
 
One way to understand the boundary condition (\ref{bound}) from the point of view of stochastic quantization is the following.
The field configuration is constrained to satisfy the Langevin evolution
(\ref{Langevin}). After infinite time the field will eventually
relax to the equilibrium configuration at the boundary $t=0$. At the
equilibrium, the role of quantum oscillation is played by the noise average. Taking the average of the Langevin equation we
directly read the boundary conditions in (\ref{bound}), which are now valid for the "quantum" configuration of the field.
The key role played by the choice of boundary conditions deserves some more analysis and in the next section we will provide a geometrical interpretation for them from the holographic point of view.

\section{Geometric interpretation and Outlook}

There is a simple, but possible far reaching geometrical origin
behind the relationship between the 4-dimensional $\phi^4$ theory and
the 3-dimensional $\phi^6$ theory. Consider the Euclidean Einstein-Hilbert action in half $\mathbb{R}_4$ 
\beq
\label{EH4d}
I_{EH}^{(4)}=\frac{-1}{16\pi G_4}\int_0^\infty dr\int d^3\vec{x}\,\sqrt{g}\left({\cal R}-2\Lambda_4\right)\,.
\eeq
It is well-known that in order to setup a proper Dirichlet problem for the metric at the boundary $r=0,\infty$, i.e. in order that the variation of bulk on-shell action vanishes when we fix the metric at the boundary, we have to add to (\ref{EH4d}) boundary the Gibbons-Hawking term \cite{GH}
\beq
\label{GH}
I_{GH}^{(4)}=\frac{1}{8\pi G_4}\int_{\partial {\cal M}} d^3\vec{x}\sqrt{g}g^{ij}K_{ij}\,,
\eeq
where $g_{ij}(\vec{x})$ is the restriction of the bulk metric to the boundary and $K_{ij}(\vec{x})$ is the extrinsic curvature. From now on we take all fields to vanish at $r=\infty$, hence the boundary $\partial{\cal M}=\mathbb{R}_3$ is at $r=0$. Consider conformally flat metrics $g_{\m\n}(x)=\varphi^2(x)\eta_{\m\n}$. An explicit calculation gives
\beq
\label{EHphi4}
I_4=I_{EH}^{(4)}+I_{GH}^{(4)}=-\frac{3}{4\pi}\int d^4x\left(\frac{1}{2}\eta^{\mu\nu}\partial_\mu\phi\partial_\nu\phi-\frac{\lambda_4}{6}\phi^4\right)\,,
\eeq
where we have defined
\beq
\label{philambda4}
\varphi(x)=\sqrt{G_4}\phi(x)\,,\,\,\,\,\lambda_4=G_4\Lambda_4=-\frac{3}{2}\lambda\,.
\eeq
The GH term cancelled exactly the boundary term arising from $I_{EH}^{(4)}$. As we have mentioned before, (\ref{EHphi4}) is equivalent to the 1st-order action (\ref{hamaction}). Hence, its on-shell variation yields
\beq
\label{deltaI4}
\delta I_4^{on-shell} = \frac{3}{4\pi}\int_{\mathbb{R}_3}d^3\vec{x}\,\delta\phi_0(\vec{x})\,\pi_0(\vec{x})\,,
\eeq
where the boundary values of the canonical variables have been defined in Section 3. 

However, if we do not wish to fix the boundary metric, the only way to make (\ref{EHphi4}) stationary is to impose (Neumann) boundary conditions on $\pi_0(\vec{x})$ e.g. the above case we should require $\pi_0(\vec{x})=0$. This gives us intriguing possibilities, for example by adding the appropriate boundary functionals we can impose boundary conditions satisfied by exact non-perturbative solutions of the bulk equations of motion. 

With this in mind we consider extending (\ref{EHphi4}) by {\it just} the 3d gravity in the boundary with action
\beq
\label{3dgravity}
I_{EH}^{(3)} = -\frac{1}{16\pi G_3}\int_{\mathbb{R}_3} d^3\vec{x}\sqrt{\hat{g}}\left({\cal R}_3-2\Lambda_3\right)\,.
\eeq
Since there are no boundary terms now, the variation of (\ref{3dgravity}) gives
\beq
\label{deltaI3}
\delta I_{EH}^{(3)}=-\frac{1}{16\pi G_3}\int d^3\vec{x}\sqrt{\hat{g}}\delta\hat{g}^{ij}\left(R_{ij}-\frac{1}{2}\hat{g}_{ij}R+\Lambda_3\hat{g}_{ij}\right)\,.
\eeq
Take now the boundary metric to be conformally flat and related to the bulk one as
\beq
\hat{g}(\vec{x})_{ij}=\varphi^2_0(\vec{x})\eta_{ij}\,,
\eeq
and using (\ref{philambda4}) we find after some algebra
\beq
\label{deltaI31}
\delta I_{EH}^{(3)}=\frac{3\Lambda_3 G_4}{8\pi}\sqrt{\frac{G_4}{G_3^2}}\int_{\mathbb{R}_3}d^3\vec{x}\,\delta\phi_0(\vec{x})\left(1-\frac{1}{6\Lambda_3G_4}{\cal R}_3\right)\phi_0^2(\vec{x})\,,
\eeq
where ${\cal R}_3$ is given by (\ref{R3}). Quite remarkably, the Hamiltonian boundary condition (\ref{HamBC}) arises as the stationarity condition for the total action
\beq
\label{I4total}
{\cal I}=I_4+I_{EH}^{(3)}\,,
\eeq
which is nothing but the sum of bulk and boundary gravity in 4d and 3d. Matching the coefficients gives the following relationships between 4d and 3d quantities
\beq
\label{4vs3d}
\lambda=-\Lambda_3G_4\,,\,\,\,\frac{2}{\lambda} G_3^2=G_4\,,\,\,\,\Lambda_3=\frac{2}{3}\Lambda_4\,,
\eeq
hence we need a negative cosmological constant in the boundary as well. 

To summarize, we presented a rather simple example of an explicit correspondence between a 4d bulk and a 3d boundary theory. We have argued that it gives support to our claim  that there is a relationship between stochastic quantization and AdS/CFT, at least under certain conditions. We pointed out that our results have a geometric origin since they can be obtained by the coupling of 4d and 3d conformal gravities.  

We need not stress again that our results must be taken only as an incentive to study further the relationship between stochastic quantization and holography \footnote{For a related but slightly different approach see \cite{Dijkgraaf}.}. In particular, one would need to understand further the subleading term in $\lambda$ and how they match between 3d and 4d, where we suspect we could find the evidence for the presence of Gaussian noise. Moreover, one could study correlation functions and extend our analysis to gauge fields and gravity. Such issues are currently under study. 

{\bf Acknowledgments:}

{\small \it The work of ACP was partially supported by the research grant with KA 2745 from the University of Crete. A. C. P. would like to thank G. Kofinas and G. Semenoff,  for useful discussions, and J. Ambjorn and P. Damgaard for interesting correspondence.  }

 \section*{Appendix}
\appendix
\section{Stochastic Quantization}

In many ways stochastic quantization may be viewed as an application of Stoke's theorem in field theory \cite{Baulieu}. 
% the idea
%of adding a new ``fictious'' time coordinate $t$. The core of the
%procedure is thus to note that ``equal-time'' correlators of the bulk
%theory tends to the correlators of the original theory living in one
%dimension less. From a geometric point of view a similar idea is
%carried on by Stokes' theorem: 
Consider a $(d+1)$-dimensional manifold
${\cal X}$ with a non empty boundary $\partial{\cal X}$. Then, for any
$d$-form $\Omega_{[d]}$ the following equality holds
\begin{equation}
\int_{{\cal X}}\dd\Omega_{[d]}=\int_{\partial{\cal X}}\Omega_{[d]}\,.\label{stokes}
\end{equation}

Now suppose that $\Omega_{[d]}$ is the lagrangian of a $d$-dimensional Euclidean theory with action $S_{cl}[\phi]$ involving a single scalar field $\phi$,
\begin{equation}
S_{cl}[\phi]=\int_{\partial{\cal X}}\Omega_{[d]}[\phi;\vec x]\,,
\end{equation}
where $\vec{x}$ are the coordinates of the boundary. The above imply that \eqref{stokes} provides an alternative definition of the $d$-dimensional theory $S_{cl}[\phi]$ using $d+1$-forms that depend on the $d+1$ coordinates $\{t,\vec{x}\}$. However,  despite the fact that we have added a new dimension - the stochastic "time" coordinate $t$ - the physical content of the theory still resides on the boundary. The latter property implies the presence of a topological invariance in the $(d+1)$-dimensional description of the theory: the action cannot depend on the specific extension of the field in the bulk. Explicitly, the $d+1$-dimensional action in terms of the field $\Phi\in{\mathscr C}^{\infty}({\cal X})$ which is the extension on ${\cal X}$ of $\phi\in{\mathscr C}^{\infty}(\partial{\cal X})$ constrained by  $\Phi\Big|_{\partial{\cal X}}=\phi$, is defined as 
\begin{equation}
{\cal S}_{d+1}[\Phi]=\int_{{\cal X}}\dd\Omega_{[d]}\,.
\end{equation}
The topological invariance can thus be phrased as ${\cal S}_{d+1}[\Phi+\delta\Phi]={\cal S}_{d+1}[\Phi]$ for any variation vanishing on the boundary $\delta\Phi\Big|_{\partial{\cal X}}=0$. The idea is to gauge-fix such a symmetry a-la BRST.

To be more concrete consider a manifold ${\cal X}$ with a cylindrical structure, ${\cal X}={\cal B}\times[0,T]$ where the base manifold ${\cal B}$ is parameterized by the coordinates $\vec{x}$ and the additional coordinate $t$ runs from $0$ to $T$. The boundary is given by the $d$-chain consisting in the two copies of ${\cal B}$ at $t=0$ and at $t=T$ with the correct orientations, $\partial{\cal X}={\cal B}_{T}-{\cal B}_{0}$. Explicitly
\begin{equation}
{\cal S}_{d+1}[\Phi]=\left(\int_{{\cal B}_{T}}-\int_{{\cal B}_{0}}\right)\Omega_{[d]}=S_{cl}[\phi_{T}]-S_{cl}[\phi_{0}]=\tilde{S}_{cl}[\phi]\,,
\end{equation}
where $\phi_{0}=\Phi\Big|_{{\cal B}_{0}}$ and $\phi_{T}=\Phi\Big|_{{\cal B}_{T}}$. For $\Omega_{[d]}$ a pure $d$-form on the base manifold we have 
\begin{equation}
\Omega_{[d]}=\dfrac1{n!}\Omega_{i_{1}\ldots i_{d}}(t,\vec x)\dd x^{i_{1}}\wedge\cdots\wedge\dd x^{i_{d}}\,,\,\,\,\,\dd\Omega_{[d]}=\dd t\wedge\dot\Omega_{[d]}\,.
\end{equation}
 But since $\Omega_{[d]}$ depends on $t$ only through the extension $\Phi$ of the scalar field we simply have that
% \begin{equation}
%\dot\Omega_{[d]}=\dot\Phi\dfrac{\delta \tilde I}{\delta\Phi}\,,
%\end{equation} 
%where $\dfrac{\delta \tilde I}{\delta\Phi}$ denotes the derivative of $\tilde I[\phi]$ with respect to $\phi$ where we substitute $\phi$ with its extension $\Phi$. Therefore
\begin{equation}
{\cal S}_{d+1}[\Phi]=\int_{{\cal X}}\dd t\wedge\dot\Omega_{[d]}=\int_{{\cal X}}\dd t\wedge\dot\Phi\dfrac{\delta S_{cl}}{\delta\Phi}\,. 
\end{equation}

%parameterization of the manifold ${\cal X}$ such that, close to the boundary, behaves like a line bundle where $\{x^{i}\}$ denote the coordinates on the boundary, being the base manifold of the bundle, and $t$ describes the direction along the fiber, towards the bulk.

Next we choose a convenient gauge-fixing condition $E[\Phi;t,\vec x]$ corresponding to a particular choice of the extension $\Phi$ of the scalar field. For example, we may consider an instantonic gauge-fixing condition given by the Langevin equation
\begin{equation}
\label{Langevin1}
E[\Phi;t,\vec x]=\dot\Phi(t,\vec x)\,\varepsilon_{[d]}(\vec x)+\kappa\dfrac{\delta S_{cl}}{\delta\Phi(t,\vec x)}\,,
\end{equation}
where $\alpha$ is a constant kernel and $\varepsilon_{[d]}$ is the volume form on the boundary. Classically we would like to take $E[\Phi;t,\vec x]=0$ but is not a good quantum condition. Hence the idea is to let it hold only as an average, $\langle E[\Phi;t,\vec x]\rangle=0$ and hence we are led to introduce a white noise $\eta$ as a source for the Langevin equation, $E[\Phi;t,\vec x]=\eta(t,\vec x)\varepsilon_{[d]}(\vec x)$. We thus introduce the ghost $\psi$ which is an anti-commuting scalar, the corresponding anti-ghost $\bar\psi$, a source $\eta$ for the gauge fixing condition which is typically given by a white noise, and a BRST nilpotent operator $Q$ such that
\begin{equation}
Q\Phi=\psi\,,\qquad Q\bar\psi=\eta,\qquad Q\psi=0\,,\qquad Q\eta=0\,.
\end{equation}
It is easy to show that the action $S_{d+1}[\Phi]$ is invariant under the fermionic transformation $\delta_{\epsilon}\equiv\bar\epsilon Q$ provided that $\psi$ satisfies periodic boundary conditions\footnote{For simplicity we will assume the fields $\psi$ and $\bar{\psi}$ to vanish on the boundary.}. We thus add a $Q$-exact term in the action which does not modify the $\delta_{\epsilon}$ invariance of the theory
\begin{equation}
{\cal S}_{d+1}[\Phi]\to {\cal S}_{d+1}[\Phi]+\left(Q S_{cl}\right)[\Phi,\eta,\psi,\bar\psi]\,,\quad{\rm with}\quad S_{cl}[\Phi,\eta,\bar\psi]=\frac{1}{2 \kappa}\int_{{\cal X}}\bar\psi\left(\eta\varepsilon_{[d]}-2E\right)\,.
\end{equation}
Doing so we arrive at
\begin{align}
\left(Q S_{cl}\right)[\Phi,\eta,\psi,\bar\psi]	&=\frac{1}{2\kappa}\int_{{\cal X}}\dd t\wedge\left[\eta\left(\eta\varepsilon_{[d]}-2E\right)+2\bar\psi\dfrac{\delta E}{\delta\Phi}\psi\right]\nonumber\\
		&=	\frac{1}{\kappa}\int_{{\cal X}}\dd t\wedge\left[\dfrac12\left(\eta-{}^{*_{d}}E\right)^{2}+\dfrac12\left({}^{*_{d}}E\right)^{2}\right]\varepsilon_{[d]}+\nonumber \\
		&	\hspace{-.3cm}
	+\frac{1}{\kappa}\iint_{{\cal X}}\dd t\wedge\bar\psi(t,\vec x)\left[\delta^{d}(\vec x-\vec y)\varepsilon_{[d]}(\vec y)\,\partial_{t}+\kappa \dfrac{\delta^{2}S_{cl}}{\delta\Phi(t,\vec x)\delta\Phi(t,\vec y)}\right]\psi(t,\vec y)\,.\label{QIaction}
\end{align}

Consider then the partition function for the theory
\begin{equation}
{\cal Z}[\phi_{0},\phi_{T}] = \int{\cal D}\Phi{\cal D}\eta{\cal D}\psi{\cal D}\bar\psi\,e^{-\left\{{\cal S}_{d+1}[\Phi]+\left(Q S_{cl}\right)[\Phi,\eta,\psi,\bar\psi]\right\}}\,.
\end{equation}
If $\kappa>0$ we can integrate out the white noise ending with the following effective action
\begin{equation}
S_{{\rm eff}}[\Phi,\psi,\bar\psi]=\int_{{\cal X}}\dd t\wedge\left[\dot\Phi\dfrac{\delta S_{cl}}{\delta\Phi}+\frac{1}{2\kappa}\left({}^{*_{d}}E\right)^{2}\varepsilon_{[d]}\right]+{\mathscr F}[\Phi,\psi,\bar\psi]\,,
\end{equation}
where ${\mathscr F}[\Phi,\psi,\bar\psi]$ is given by the last line of \eqref{QIaction}. Then, using
\begin{equation}
\left({}^{*_{d}}E\right)^{2}=\dot\Phi^{2}+ 2\kappa\dot\Phi\;{}^{*_{d}}\dfrac{\delta S_{cl}}{\delta\Phi}+ \kappa^{2} \left({}^{*_{d}}\dfrac{\delta S_{cl}}{\delta\Phi}\right)^{2}\,,
\end{equation}
we obtain
\begin{equation}
S_{{\rm eff}}[\Phi,\psi,\bar\psi]=\int_{{\cal X}}\dd t\wedge\left[\frac{1}{2\kappa}\dot\Phi^{2}\varepsilon_{[d]}+ 2 \dot\Phi\dfrac{\delta S_{cl}}{\delta\Phi} +\frac{\kappa}{2} \left({}^{*_{d}}\dfrac{\delta\ S_{cl}}{\delta\Phi}\right)^{2}\varepsilon_{[d]}\right]+{\mathscr F}[\Phi,\psi,\bar\psi]\,.
\end{equation}
The second term is a total derivative contribution which vanishes if we choose periodic boundary conditions for $\Phi$. We thus have
\begin{equation}
S_{{\rm eff}}[\Phi,\psi,\bar\psi]=\frac{1}{\kappa}\int_{{\cal X}}\dd t\wedge\left[\dfrac12\dot\Phi^{2}\varepsilon_{[d]}+\frac{\kappa^{2}}{2}\left({}^{*_{d}}\dfrac{\delta S_{cl}}{\delta\Phi}\right)^{2}\varepsilon_{[d]}\right]+{\mathscr F}[\Phi,\psi,\bar\psi]\,.
\end{equation}
The fermonic part of the action can be formally integrated out to give
\begin{equation}
\int{\cal D}\psi{\cal D}\bar\psi\,e^{-{\mathscr F}[\Phi,\psi,\bar\psi]}={\rm det}\left(\dfrac{\delta E}{\delta\Phi}\right) \sim \exp\left[\kappa \int_{{\cal X}}\dd t\wedge\dfrac{\delta^{2} S_{cl}}{\delta\Phi(t,\vec x)^{2}}\right] - \exp\left[-\kappa\int_{{\cal X}}\dd t\wedge\dfrac{\delta^{2} S_{cl}}{\delta\Phi(t,\vec x)^{2}}\right]\,.
\end{equation}
In this step one must be consistent with the choice of periodicity for the fields on the boundary.
Collapsing the forms we can then write the end result as
\begin{equation}
\label{FP1}
{\cal Z} = \int{\cal D}\Phi \left[ e^{-S^+_{FP}}-e^{-S^-_{FP}} \right], \qquad S^{\pm}_{FP}[\Phi]= \int d^{d+1}x\left[\frac{1}{2\kappa}\dot{\Phi}^2+\frac{\kappa}{2} \left(\frac{\delta S_{cl}}{\delta\Phi}\right)^2 \pm \kappa \frac{\delta^2 S_{cl}}{\delta\Phi^2}\right]\,.
\end{equation}
It's important to stress that this derivation of the partition function for the FP system leads to the supersymmetric realization of stochastic quantization \cite{Gozzi}. Supersymmetry follows from the choice of periodic boundary conditions for the fields $\psi$ and $\Phi$. It's straightforward to reintroduce in the partition function the dependence on general choices of boundary values for the fields and consider only forward propagation in time. In this more general setting one is free to fix a given initial configuration for the fields and let them evolve according to the Langevin equation. In Section 2 we explicitly chose an initial condition for the field $\phi$  generally breaking supersymmetry.


\begin{thebibliography}{99}

\bibitem{DPolyakov}
  D.~Polyakov,
  %``AdS/CFT correspondence, critical strings and stochastic quantization,''
  Class.\ Quant.\ Grav.\  {\bf 18}, 1979 (2001)
  [arXiv:hep-th/0005094].
  %%CITATION = CQGRD,18,1979;%%

\bibitem{Periwal}
  G.~Lifschytz and V.~Periwal,
  %``Schwinger-Dyson = Wheeler-DeWitt: Gauge theory observables as bulk
  %operators,''
  JHEP {\bf 0004} (2000) 026
  [arXiv:hep-th/0003179].
  %%CITATION = JHEPA,0004,026;%%

\bibitem{ParisiWu}
  G.~Parisi and Y.~s.~Wu,
  %``Perturbation Theory Without Gauge Fixing,''
  Sci.\ Sin.\  {\bf 24}, 483 (1981).
  %%CITATION = SSINA,24,483;%%

\bibitem{AmbjornLoll}
  J.~Ambjorn, R.~Loll, W.~Westra and S.~Zohren,
  %``Stochastic quantization and the role of time in quantum gravity,''
  Phys.\ Lett.\  B {\bf 680}, 359 (2009)
  [arXiv:0908.4224 [hep-th]].
  %%CITATION = PHLTA,B680,359;%%

\bibitem{Damgaard}
  P.~H.~Damgaard and H.~Huffel,
  %``Stochastic Quantization,''
  Phys.\ Rept.\  {\bf 152}, 227 (1987).
  %%CITATION = PRPLC,152,227;%%

\bibitem{Zinn-Justin}
  J.~Zinn-Justin,
  %``RENORMALIZATION AND STOCHASTIC QUANTIZATION,''
  Nucl.\ Phys.\  B {\bf 275} (1986) 135.
  %%CITATION = NUPHA,B275,135;%%

\bibitem{Horava}
  P.~Horava,
  %``Quantum Gravity at a Lifshitz Point,''
  Phys.\ Rev.\  D {\bf 79} (2009) 084008
  [arXiv:0901.3775 [hep-th]].
  %%CITATION = PHRVA,D79,084008;%%

\bibitem{Marolf}
  G.~Compere and D.~Marolf,
  %``Setting the boundary free in AdS/CFT,''
  Class.\ Quant.\ Grav.\  {\bf 25}, 195014 (2008)
  [arXiv:0805.1902 [hep-th]].
  %%CITATION = CQGRD,25,195014;%%

\bibitem{dHP}
  S.~de Haro and A.~C.~Petkou,
  %``Instantons and conformal holography,''
  JHEP {\bf 0612} (2006) 076
  [arXiv:hep-th/0606276].
  %%CITATION = JHEPA,0612,076;%%

\bibitem{dHPP}
  S.~de Haro, I.~Papadimitriou and A.~C.~Petkou,
  %``Conformally coupled scalars, instantons and vacuum instability in
  %AdS(4),''
  Phys.\ Rev.\ Lett.\  {\bf 98} (2007) 231601
  [arXiv:hep-th/0611315].
  %%CITATION = PRLTA,98,231601;%%

\bibitem{Papadimitriou}
 I.~Papadimitriou,
  %``Multi-Trace Deformations in AdS/CFT: Exploring the Vacuum Structure of
  %the Deformed CFT,''
  JHEP {\bf 0705} (2007) 075
  [arXiv:hep-th/0703152].
  %%CITATION = JHEPA,0705,075;%%

\bibitem{wittenAdSCFT}
  E.~Witten,
  %``Anti-de Sitter space and holography,''
  Adv.\ Theor.\ Math.\ Phys.\  {\bf 2} (1998) 253
  [arXiv:hep-th/9802150].
  %%CITATION = 00203,2,253;%%

\bibitem{Salopek}
  J.~Parry, D.~S.~Salopek and J.~M.~Stewart,
  %``Solving the Hamilton-Jacobi equation for general relativity,''
  Phys.\ Rev.\  D {\bf 49} (1994) 2872
  [arXiv:gr-qc/9310020].
  %%CITATION = PHRVA,D49,2872;%%

\bibitem{GH}
  G.~W.~Gibbons and S.~W.~Hawking,
  %``Action Integrals And Partition Functions In Quantum Gravity,''
  Phys.\ Rev.\  D {\bf 15} (1977) 2752.
  %%CITATION = PHRVA,D15,2752;%%
  
\bibitem{Baulieu}
  L.~Baulieu and B.~Grossman,
  %``A TOPOLOGICAL INTERPRETATION OF STOCHASTIC QUANTIZATION,''
  Phys.\ Lett.\  B {\bf 212} (1988) 351.
  %%CITATION = PHLTA,B212,351;%%

\bibitem{Gozzi}
  E.~Gozzi,
  %``Functional Integral Approach To Parisi-Wu Stochastic Quantization: Scalar
  %Theory,''
  Phys.\ Rev.\  D {\bf 28} (1983) 1922.
  %%CITATION = PHRVA,D28,1922;%%

\bibitem{Dijkgraaf}
  R.~Dijkgraaf, D.~Orlando and S.~Reffert,
  %``Relating Field Theories via Stochastic Quantization,''
  Nucl.\ Phys.\  B {\bf 824} (2010) 365
  [arXiv:0903.0732 [hep-th]].
  %%CITATION = NUPHA,B824,365;%%

\end{thebibliography}
\end{document}